\documentclass[prb,superscriptaddress,showpacs,twocolumn]{revtex4}%
\usepackage{graphicx}
\usepackage{amsmath}
\usepackage{amsfonts}
\usepackage{color}
\usepackage{amssymb}%
\begin{document}
\author{Guoren Zhang}
\affiliation{
Institute for Advanced Simulation, Forschungszentrum J\"ulich, 
D-52425 J\"ulich, Germany}
\author{Evgeny Gorelov}
\affiliation{
Institute for Advanced Simulation, Forschungszentrum J\"ulich, 
D-52425 J\"ulich, Germany}
\author{Erik Koch}
\affiliation{German School for Simulation Sciences, 52425 J\"ulich, Germany}
\affiliation{JARA High-Performance Computing}
\author{Eva Pavarini}
\affiliation{
Institute for Advanced Simulation, Forschungszentrum J\"ulich, 
D-52425 J\"ulich, Germany}
\affiliation{JARA High-Performance Computing}
\title{Importance of exchange-anisotropy and superexchange for the\\ spin-state transitions in LnCoO$_3$ (Ln=La,Y,RE) cobaltates}
\begin{abstract}
Spin-state transitions are the hallmark of rare-earth cobaltates. In order to understand them, it is essential to identify all relevant parameters which shift the energy balance between spin states, and determine their trends. We find that $\Delta$, the
 $e_g$-$t_{2g}$ crystal-field splitting, increases by $\sim$ 250 meV when increasing pressure to 8 GPa and by about 150~meV when cooling from 1000~K to 5~K. It changes, however, by less than 100 meV when La is substituted with another rare earth.
Also the Hund's rule coupling ${\cal{J}}_{\rm avg}$ is about the same in systems with very different spin-state transition temperature, like LaCoO$_3$ and EuCoO$_3$. 
Consequently, in addition to $\Delta$ and ${\cal{J}}_{\rm avg}$, the Coulomb-exchange anisotropy $\Delta {\cal{J}}_{\rm avg}$ and the super-exchange energy-gain $\Delta E_{\rm SE}$  play a crucial role, and are comparable with spin-state dependent relaxation effects due to covalency. 
We show that in the LnCoO$_3$ series, with Ln=Y or a rare earth (RE), super-exchange progressively stabilizes a low-spin ground state as the Ln$^{3+}$ ionic radius decreases.
We give a simple model to describe spin-state transitions and show that, at low temperature, the formation of isolated high-spin/low-spin pairs is favored, while in the high-temperature phase, the most likely homogeneous state is high-spin, rather than intermediate spin. An {\em orbital-selective} Mott state could be a fingerprint of such a state.
\end{abstract}
\maketitle

\section{Introduction}
\begin{figure}
\center
\rotatebox {0}{\includegraphics[width=0.43\textwidth]{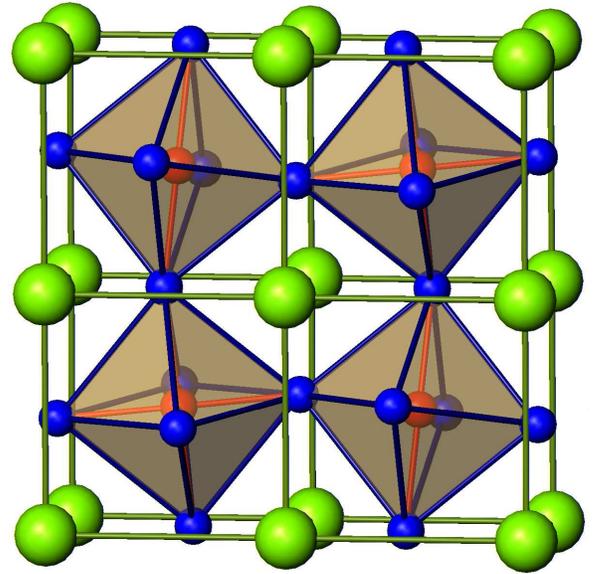}}
\caption{(Color on-line) \label{crys-str}
Rhomobohedral structure of LaCoO$_3$. 
We define the ${\bf x}$, ${\bf y}$, ${\bf z}$ pseudo-cubic axes as those connecting the Co atoms; 
 ${\bf z}$ is the vertical axis, ${\bf y}$ the horizontal axis, and ${\bf x}$ is perpendicular to the plane of the figure.
 }
\end{figure}

The nature of the spin-state transitions in  LaCoO$_3$  (Fig.~\ref{crys-str})  is the subject of controversy for decades.\cite{goodenough,bhide,haverkort,podlesnyak,sundaram,gaps,asai,yamaguchi,potze,saitoh,noguchi,korotin,klie,zobel,Ishikawa,asai2,knizek2,zhuang,knizek1,laref}
LaCoO$_3$  undergoes two electronic crossovers, at $T_{\rm SS}\sim 50-100~$K and  
$T_{\rm IM}\sim500-600$~K; the first is commonly ascribed to a change in spin-state, while the second to an insulator to metal transition.\cite{bhide,gaps} There is a general agreement that
the ground state of LaCoO$_3$ is insulating and non-magnetic, with Co in the low-spin (LS) $t_{2g}^6$ state. 
The core of the debate is whether the $t_{2g}^4e_g^2$ high-spin (HS) state\cite{goodenough,bhide,haverkort,podlesnyak,sundaram}  
($S\!=\!2$), or the $t_{2g}^5e_g^1$ intermediate-spin (IS) state\cite{potze,saitoh,noguchi,korotin,klie,zobel,Ishikawa} ($S\!=\!1$) is thermally excited right above $T_{\rm SS}$, and if the spin-state crossover involves two or more steps between  $T_{\rm SS}$ and $T_{\rm IM}$. 
Various mixed phases\cite{asai2,knizek2} and spin-state superlattices\cite{zhuang,knizek1,laref,Kunes} have  been suggested as final or intermediate step. For the high-temperature phase, a pure IS state has been proposed based on LDA+$U$ results\cite{knizek2} or phenomenological thermodynamic analysis.\cite{kiomen} 
To date, the controversy  remains open.

Spin-state transitions have also been reported in LaCoO$_3$ under pressure,\cite{pressure1,pressure2,pressure3,pressure4}
as well as in several LnCoO$_3$ perovskites, where Ln is Y or a rare earth (RE) heavier than La.\cite{ystr,recoo1,recoo2}
In the LnCoO$_3$ series, the spin-state and, to a smaller extent, the insulator to metal
transition, shifts to higher temperatures with decreasing RE$^{3+}$ ionic radius.\cite{ystr,recoo1,recoo2,recoo3}
Since HS and IS states are Jahn-Teller (JT) active, one might expect co-operative JT distortions in correspondence with the spin-state transition. Remarkably, 
early neutrons diffraction data\cite{str1,str2,str3} indicate that the structure of LaCoO$_3$ is rhombohedral (space group $R\bar3c$) with no co-operative JT distortion at all,\cite{str1,str2,str3,sundaram} while more recently a co-operative JT distortion (space group $I2/a$) has been found via single-crystal X-ray diffraction.\cite{maris,zobel} 
The presence of local JT distortions  is as controversial.\cite{louca,sundaram}
On the other hand, LnCoO$_3$ perovskites with Ln$\ne$La are orthorhombic (space group $Pbnm$) and do exhibit a small, weakly temperature dependent,
JT distortion.\cite{ystr,recoo1,recoo3}

The  LS$\to$IS scenario\cite{potze} was popularized by total-energy LDA+$U$ calculations,\cite{korotin} which show that for rhombohedral LaCoO$_3$ the homogeneous IS state is almost degenerate with the LS ground state, and sizably lower in energy than the  HS state.\cite{korotin,zhuang,knizek1,knizek2} 
The LS$\to$IS scenario  has  been used to interpret experiments pointing to a spin triplet.\cite{noguchi,recoo1,zobel} 
However, a triplet can also arise from the splitting of the HS state via spin-orbit interaction.\cite{radwanski,klie,pandey} 
Furthermore, it has been shown with LDA+$U$ and unrestricted Hartree-Fock that some inhomogeneous LS/HS phases and/or superlattices,\cite{zhuang,knizek1,knizek2} are more stable than the homogeneous IS state. 
Experimentally, neither the HS+LS nor other superlattices have been reported so far; instead, there are strong indications that X-ray absorption spectroscopy (XAS) and magnetic circular dichroism (MCD) data\cite{haverkort} as well as inelastic neutron scattering experiments\cite{podlesnyak} are compatible with a mixed HS/LS phase. It has been pointed out that the lattice could play a crucial role in such mixed phases, by expanding around HS ions due to covalency effects\cite{goodenough,kiomen,haverkort} 

While the experimental evidence in favor of  LS $\to$ HS crossover at $T_{\rm SS}$, perhaps with disorder phenomena, is growing,\cite{noguchi,podlesnyak,haverkort} the actual parameters which tip the energy balance in favor of a given spin-state in the real materials have not been fully identified, and, to the best of our knowledge no systematic attempt to determine their evolution in the LnCoO$_3$ series exist. 
Furthermore, the interplay between spin-state and insulator to metal transition, even in the simplest homogeneous scenarios, and the nature of the high-temperature metallic phase, are not fully understood.

In the present work, we study the electronic structure trends of the LnCoO$_3$ family and identify
the material-dependent parameters responsible for the spin-state transitions. 
We show that,  for realistic Coulomb and exchange parameters,  an IS homogeneous phase is very unlikely in any of the considered materials, both at low and high temperature. We show that, besides the interplay of the $e_g$-$t_{2g}$ crystal-field $\Delta$ and  the Coulomb exchange ${\cal{J}}_{\rm avg}$,
the exchange anisotropy $\Delta {\cal{J}}_{\rm avg}$ and the super-exchange energy gain $\Delta_{\rm SE}$
 stabilize the LS ground state. 
Our results support a scenario in which HS-LS pairs, the formation of which is favored by both super-exchange interaction and covalency effects, are excited at $T_{\rm SS}$. We find that,  for all known experimental structures, the Jahn-Teller crystal-field splitting is weak.
In the high-temperature regime,  we find, within a homogeneous scenario, a HS state; in such a state, the insulator-to-metal transition  depends mostly on the $t_{2g}$ degrees of freedom.
 Our results could be used to distinguish a pure high-temperature HS state from a mixed HS/LS state.

\section{Method}
We calculate the electronic structure using two {\it ab-initio} approaches based on density-functional theory
in the local-density approximation (LDA). The first is the Linear Augmented Plane Wave (LAPW) method as implemented in the Wien2k code;\cite{wien2k} to obtain hopping integrals and crystal-field splitting we use maximally localized Wannier functions.\cite{mlwf} The second approach is the downfolding technique based on the $N$th-Order Muffin-Tin Orbital (NMTO) method in the form discussed in Refs.~\onlinecite{evad1}.  
By means of these two methods, we construct the generalized Hubbard model for the Co $d$ bands
\begin{eqnarray}\label{H}
\nonumber
H&=&-\!\sum_{im,i^\prime m^\prime \sigma} t_{m,m^\prime}^{i,i^\prime}
c^{\dagger}_{im\sigma} c^{\phantom{\dagger}}_{i^\prime m^\prime\sigma}
\\
&+&\!
\frac{1}{2}\sum_{i\sigma,\sigma^\prime} 
\sum_{m m^\prime}
\sum_{p p^{\prime}} 
U_{m p m^{\prime} p^{\prime}} 
c^{\dagger}_{im\sigma}           c^{\dagger}_{ip\sigma^\prime}
c^{\phantom{\dagger}}_{ip^\prime\sigma^\prime} c^{\phantom{\dagger}}_{im^\prime\sigma}.
\end{eqnarray}
Here $c^{\dagger}_{im\sigma}$ creates an electron with spin $\sigma$  in the Wannier orbital $m$
at site $i$ ($m=xz,yz,xy,x^2-y^2,3z^2-r^2$). $U_{m p m^{\prime} p^{\prime}} $ are 
rotationally invariant screened Coulomb integrals.
 The parameters $U_{m p m^{\prime} p^{\prime}}$ 
can all be expressed as a function of the (screened) Slater integrals $F_0$, $F_2$ and $F_4$.
We adopt the common definition for the average direct and exchange couplings,
$U_{\rm avg}=F_0$, and $J_{\rm avg}=\frac{1}{14} (F_2+F_4)$.
A pedagogical derivation of the rotationally invariant Coulomb matrix  can
be found, e.g., in Ref.~\onlinecite{lda+dmft}.
Apart from the average exchange interaction, the exchange anisotropy plays a significant role.
It is therefore convenient to express all parameters as linear combinations of $U_{\rm avg}$, ${\cal{J}}_{\rm avg}=\frac{5}{7} J_{\rm avg}$, and 
$\Delta {\cal{J}}_{\rm avg}={\cal{J}}_{\rm avg} ({\frac{1}{5} - \frac{1}{9} \frac{F_4}{F_2}})/({1+\frac{F_4}{F_2}})$; the latter measures the anisotropy in the exchange interactions.
The exchange parameters for the $t_{2g}$ and $e_g$ states are then $J_{t_{2g}}={\cal{J}_{\rm avg}}+\Delta {\cal{J}_{\rm avg}}$ and  %
 $J_{e_g}={\cal{J}}_{\rm avg}+3\Delta{\cal{J}}_{\rm avg}$, respectively; 
 the Coulomb exchange parameters  between $e_g$ and $t_{2g}$ states are
${J}_{3z^2-r^2,xz}= {J}_{3z^2-r^2,yz}={\cal{J}}_{\rm avg}-3\Delta{\cal{J}}_{\rm avg}$,
${J}_{x^2-y^2,xy}={\cal{J}}_{\rm avg}-5\Delta{\cal{J}}_{\rm avg}$, ${J}_{3z^2-r^2,xy}=J_{e_g}$, and
${J}_{x^2-y^2,xz}={J}_{x^2-y^2,yz}=J_{t_{2g}}$; the orbital-diagonal direct exchange is $U_0=U_{\rm avg}+\frac{8}{5} {\cal{J}}_{\rm avg}$.
For some materials, we calculate the screened Coulomb integrals $U_{\rm avg}$ and
${{\cal J}}_{\rm avg}$ using the constrained local-density approximation (cLDA) approach.\cite{cLDA}

We solve Hamiltonian (\ref{H}) using different approaches.
In section III.A  we present exact diagonalization results (atomic limit).
In section III.B we show results obtained with second-order perturbation theory (super-exchange energy gain).
In section III.C we present dynamical mean-field theory  calculations within the LDA+DMFT (local-density approximation + dynamical mean-field theory) approach; for the latter we use a weak-coupling continuous-time quantum Monte Carlo\cite{wc} (CT-QMC) 
quantum-impurity solver and work with the full self-energy matrix in orbital space,\cite{evad1,OO}
as discussed in Ref.~\onlinecite{gorelov}. 
We obtain the spectral-function matrix by means of stochastic reconstruction.\cite{mishchenko}
Since LDA+DMFT calculations for a 5-band model with full Coulomb vertex\cite{almostfull} and self-energy matrix are computationally very expensive, a massively-parallel implementation as presented in Ref.~\onlinecite{gorelov} 
is essential.
Structural data for LaCoO$_3$ are taken from Refs.~\onlinecite{str2,maris}, for YCoO$_3$ from Refs.~\onlinecite{ystr},
and for the other LnCoO$_3$ materials\cite{noteonfelectrons} from Refs.~\onlinecite{restr1,recoo2,recoo1}.
High-pressure data are from Ref.~\onlinecite{pressure1}.

\section{Results}
\subsection{Atomic limit, Jahn-Teller, and constraints}
\begin{figure}\label{atomic}
\center
\rotatebox {270}{\includegraphics [width=0.28\textwidth]{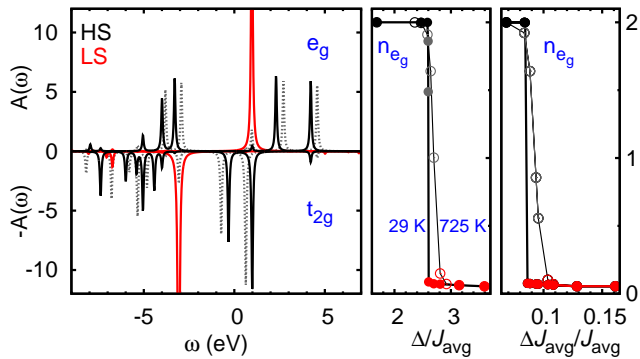}}
\caption{(Color on-line) \label{atomic}
Left: Atomic spectral-functions at 29~K for different values of the ratio
 $\Delta/{\cal{J}}_{\rm avg}$.
Full line, dark:  $\Delta/{\cal{J}}_{\rm avg} \sim 1.69$.  
Dotted line:      $\Delta/{\cal{J}}_{\rm avg} \sim 2.60$. 
Full line, light: $\Delta/{\cal{J}}_{\rm avg} \sim 2.61$.  
Coulomb parameters: $U_{\rm avg}=3~$eV, ${\cal{J}}_{\rm avg}=0.89$~eV. Center: occupation of
the $e_g$ states as a function of  $\Delta/{\cal{J}}_{\rm avg}$ at low and high temperature.
Right: occupation of the $e_g$ states as a function of  $\Delta{\cal{J}}_{\rm avg}/{\cal{J}}_{\rm avg}$ at low and high temperature; the results are obtained for ${\cal{J}}_{\rm avg}\sim 0.89$~eV, $\Delta/{\cal{J}}_{\rm avg} \sim 2.64$ and varying $F_4/F_2$ in the interval $[0.1,0.8]$.
}
\end{figure}

Let us examine the eigenstates of (\ref{H}) in the atomic limit,  i.e., in the limit $t_{m,m^\prime}^{i,i^\prime}=0$ for $i\ne i^\prime$ (see Fig.~\ref{atomic}).
The crystal-field states, the states which diagonalize the on-site matrix
$t_{m,m^\prime}^{i,i}$, have energy $\varepsilon_\alpha$, with $\varepsilon_{\alpha}\le\varepsilon_{\alpha+1}$;
$\alpha=1,2,3$ are $t_{2g}$-like states and $\alpha=4,5$ $e_g$-like.
In the atomic limit, the energy of the low-spin state ($t_{2g}^6$) is 
\begin{eqnarray*}
E_L \sim 15 U_0-30 ({\cal{J}}_{\rm avg}+\Delta {\cal{J}}_{\rm avg}) +2\sum_{\alpha\in t_{2g}}\varepsilon_\alpha.
\end{eqnarray*} 
The states with intermediate spin ($t_{2g}^5e_g^1$) can be written as 
$|\underline{\alpha},\beta\rangle$, where ${\alpha}$ is the $t_{2g}$  hole orbital 
and $\beta$ the $e_g$ electron orbital.
The energy of the IS, in the limit in which only the density-density Coulomb terms contribute, is
\begin{eqnarray*}
E_I %
    &\sim& E_L-3{\cal{J}}_{\rm avg}+ (23+f)\Delta {\cal{J}}_{\rm avg}+\Delta^{\rm LI},
\end{eqnarray*} 
where $\Delta^{\rm LI}=\varepsilon_4-\varepsilon_3$; the factor $f$ yields the deviation from the average anisotropy of the exchange interaction; $f=-8$ for $|\underline{xy},x^2-y^2\rangle$ and cyclic $xyz$ permutations.
The energy of a high-spin $t_{2g}^4e_g^2$ is
\begin{eqnarray*}
E_H %
&\sim&  E_L-8{\cal{J}}_{\rm avg}+30\Delta {\cal{J}}_{\rm avg}+2\Delta^{\rm LH}, 
\end{eqnarray*} 
where $\Delta^{\rm LH}=(\varepsilon_5+\varepsilon_4-\varepsilon_3-\varepsilon_2)/2$. 
From energy differences between spin configurations we can obtain constraints for the parameters.
Theoretical estimates\cite{Eder} yield $F_2\sim10.64$~eV and $F_4\sim 6.8$~eV; 
with these values 
${\cal{J}}_{\rm avg}\sim 0.89$~eV, and
$\Delta{\cal{J}}_{\rm avg}\sim 0.07$~eV $\sim 0.08{\cal{J}}_{\rm avg}$;
other estimates\cite{MF96} yield slightly smaller values;
by using cLDA we obtain ${\cal{J}}_{\rm avg}\sim 0.7$~eV;
furthermore, we find that ${\cal{J}}_{\rm avg}$ varies little when La is replaced by Eu in EuCoO$_3$,  a material which apparently {exhibits no spin-state transition below the insulator-to-metal transition}.\cite{recoo2} 
Since screened Coulomb exchange parameters can only be obtained within given approximations, we will discuss our results for a range of plausible
values of ${\cal{J}}_{\rm avg}$, ranging from 0.5~eV to 1~eV.
\begin{figure}
\center
\rotatebox {0}{\includegraphics [width=0.47\textwidth]{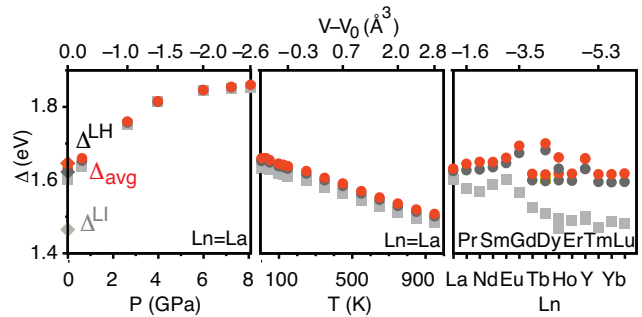}}
\caption{\label{delta}
(Color on-line) Crystal-field splittings as a function of pressure (left), temperature (center), and of Ln ion (right). The corresponding volumes are also given; $V_0$ is the volume of LaCoO$_3$ at room temperature and ambient pressure.
$\Delta^{\rm LI}=\varepsilon_4-\varepsilon_3$ (squares), $\Delta_{\rm avg}=(\varepsilon_4+\varepsilon_5)/2
-(\varepsilon_1+\varepsilon_2+\varepsilon_3)/3$ (dark circles),  and 
$\Delta^{\rm LH}=(\varepsilon_4+\varepsilon_5)/2-(\varepsilon_2+\varepsilon_3)/2$ (light circles) are shown.
In the cubic limit, $\Delta^{\rm LI}=\Delta^{\rm LH}=\Delta_{\rm avg}=\Delta$.
Multiple points for the same compound are results from different structural data sets.}
\end{figure}

In Fig.~\ref{delta} we show the calculated crystal-field splitting. 
We calculate the splittings in LaCoO$_3$  at ambient pressure and for increasing
temperature,\cite{str1,str2} at room temperature for increasing pressure\cite{pressure1,pressure2,pressure3,pressure4,lisivo} and in the series of RECoO$_3$ compounds.\cite{ystr,restr1,restr2,restr3}
Our results show that  $\Delta ^{\rm LH}\sim 1.6-1.7$~eV in the full series,\cite{nmtocef}
while  $\Delta ^{\rm LI}\sim 1.4-1.6$~eV.
From these results we can derive the following conclusions.
The ground state is LS if these conditions are met 
\begin{eqnarray*} 
\Delta ^{\rm LH} &\gtrsim& 4{\cal{J}}_{\rm avg} -15\Delta {\cal{J}}_{\rm avg} \sim 2.8  \,{\cal{J}}_{\rm avg}, \\
\Delta^{\rm LI} &\gtrsim& 3{\cal{J}}_{\rm avg} -(23+f)\Delta{\cal{J}}_{\rm avg} \sim  1.16\mbox{\rm -}1.8\,{\cal{J}}_{\rm avg},
\end{eqnarray*}
where for the second condition we give the range varying $f$ from 0 (average) to $-8$ ($|\underline{xy},x^2-y^2\rangle$).
Let us consider the cubic case ($\Delta ^{\rm LH}=\Delta ^{\rm LI}=\Delta_{\rm avg}=\Delta$).
In this case, the first condition is the most stringent. Exact diagonalization of the atomic limit Hamiltonian with full exchange interaction and  $ \Delta{\cal{J}}_{\rm avg}/{\cal{J}}_{\rm avg}\sim 0.08$ (Fig.~\ref{atomic}) indeed leads to a switch between a high and low spin ground state  at slightly smaller values, $\Delta  \sim 2.6 \,{\cal{J}}_{\rm avg}$.
From these results we conclude that, to explain a $S=0$ ground state in the atomic limit, ${\cal{J}}_{\rm avg}$ must be slightly smaller than estimated with cLDA and similar approaches, ${\cal{J}}_{\rm avg}\sim 0.6$~eV; alternatively, less likely, ${\cal{J}}_{\rm avg}\sim 0.7$~eV but the anisotropy should be sizably larger than in the free atom,\cite{atomic} $\Delta {\cal{J}}_{\rm avg}\sim 0.12{\cal{J}}_{\rm avg}$.
In any case, with the crystal field splittings in Fig.~\ref{delta}, we are quite close to the HS/LS crossover.
In this parameter region, the $\sim 100$-$200$~meV changes in crystal-field that we find (Fig.~\ref{delta}) with increasing temperature or pressure
could indeed explain {\em alone} a spin-state crossover from LS to HS or viceversa.
For a IS ground state, the following conditions should be met
\begin{eqnarray*}
2\Delta^{\rm LH}-\Delta^{\rm LI} &\gtrsim &\left(5-(7-f) \frac{ \Delta {\cal{J}}_{\rm avg} } { {\cal{J}}_{\rm avg} }
\right){\cal{J}}_{\rm avg}\sim 4.5\mbox{\rm -}3.8{\cal{J}}_{\rm avg}\\
\Delta^{\rm LI} &\le& 3{\cal{J}}_{\rm avg} -(23+f)\Delta{\cal{J}}_{\rm avg} \sim 1.16\mbox{\rm -}1.8 \,{\cal{J}}_{\rm avg}
\end{eqnarray*}
This definitely excludes a IS ground state in the cubic case as, in all systems considered, the two conditions are never satisfied at the same time. For a LS$\to$ IS scenario, it could be however sufficient that  the IS state is lower than the HS (first condition). 
In the cubic case, for the parameters in Fig.~\ref{delta}, this could happen only for apparently unrealistically small Coulomb exchange ${\cal{J}}_{\rm avg}$ ($\sim 0.4$~eV or smaller) 
or unrealistically large\cite{atomic}  anisotropy $\Delta {\cal{J}}_{\rm avg}  / { {\cal{J}}_{\rm avg} }$ ($\sim 0.16$); in these cases the IS would, however, be quite high in energy ($\sim$1~eV or more) above the LS.
Thus  the predictions of conventional ligand-field theory\cite{lft} are in agreement with our results
for the atomic limit.

Let us analyze the effects of crystal distortions on the crystal-field splittings
of $e_g$  ($\Delta_{54}=\varepsilon_5-\varepsilon_4$) and $t_{2g}$ 
($\Delta_{31}=\varepsilon_3-\varepsilon_1$, $\Delta_{21}=\varepsilon_2-\varepsilon_1$) states,
and the corresponding crystal-field orbitals. 
\begin{figure}[t]
\center
\rotatebox {0}{\includegraphics [width=0.48\textwidth]{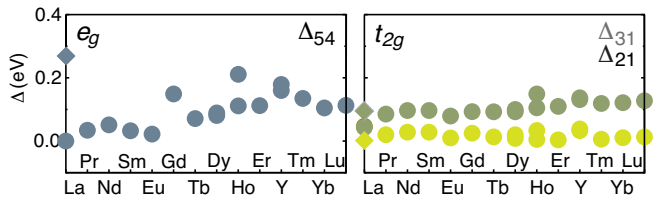}}
\caption{ \label{JT}
(Color on-line) Crystal-field splitting (in eV) of $e_g$ (left) and $t_{2g}$ (right) states at room temperature as a function of decreasing ionic radius. Multiple points for the same compound are results from different structural data.~\cite{ystr,restr,restr1,restr2,restr3}
LaCoO$_3$: $I2/a$ (diamonds) and $R\bar{3}c$ (circles).}
\end{figure}
In Fig.~\ref{JT} we show $\Delta_{54}$, $\Delta_{31}$ and $\Delta_{21}$ along the series; we find that the splitting of $e_g$ states, $\Delta_{54}$, reaches
a maximum of 200~meV in HoCoO$_3$, and 250~meV in tetragonal\cite{maris} LaCoO$_3$,
and is zero in rhombohedral LaCoO$_3$;
furthermore, we find that $\Delta_{21}$ is small while $\Delta_{54}$ and $\Delta_{31}$ are comparable, the only exception being the $I2/a$ LaCoO$_3$ structure; finally, across the spin-state transition, for available structural data, we find either basically no change in $e_g$ and $t_{2g}$ crystal-field splittings (LaCoO$_3$, $R\bar{3}c$) or a tiny change of 40~meV and no significant change in occupied orbital (YCoO$_3$). 
Even if we assume that the crystal-field splittings in Fig.~\ref{JT} are completely ascribed to the Jahn-Teller effect, they are small compared to Jahn-Teller systems such as manganites or cuprates;\cite{OO} 
they are comparable with splittings arising from the GdFeO$_3$-type distortion ($Pbnm$ structures)\cite{evad1,evad2} which progressively increase along the LnCoO$_3$ series.\cite{restr2,restr3} 
LaCoO$_3$  has to be discussed separately. 
While the presence of JT distortion is still controversial experimentally, 
even the monoclinic structure with co-operative Jahn-Teller distortion reported in Ref.~\onlinecite{maris} leads to a small splitting of $e_g$ and $t_{2g}$ states (see Fig.~\ref{JT}), although larger than for all other LnCoO$_3$ compounds considered here. 

The splittings of $e_g$ and $t_{2g}$ states might affect the energy balance between spin states.  To quantify this effect,
we introduce the average splitting ${\delta}\Delta=(\Delta_{54}+\Delta_{31})/2$. Then, on the basis of Fig.~\ref{JT},
$\Delta^{\rm LH}\sim\Delta$, $\Delta^{\rm LI} \sim\Delta-{\delta}\Delta$.
In the light of presently known structural data,  remarkably, in the full LnCoO$_3$ series, 
${\delta}\Delta$ is at most 150~meV, which is insufficient to lead to a LS-IS scenario, which would require
$\delta\Delta\sim 1\mbox{\rm -}1.7 \; {\cal J}_{\rm avg}$ or larger. 
Such a scenario could perhaps start to play a role
under high pressure, if ${\delta}\Delta$ sizably increases.

Our results show that the anisotropy $\Delta {\cal{J}}_{\rm avg}$ plays an important role in stabilizing the low-spin ground state (Fig.~\ref{atomic}); 
if we neglect it ($J_1={\cal{J}}_{\rm avg}={\cal{J}}$), an approximation
often adopted, the constraint for a low spin ground state becomes $\Delta^{\rm LH} > 4 {\cal{J}}$. 
Thus, if the $e_g$-$t_{2g}$ crystal-field splitting $\Delta$ has the values of  $\Delta_{\rm avg}$ in Fig.~\ref{delta}, the effective Coulomb exchange has to be as small as ${\cal{J}}\sim 0.4$~eV in order to obtain a LS ground state. On the other hand, if the anisotropy is  unrealistically larger than for atomic orbitals, the IS state can become the first excited state.
As a consequence, for systems close to spin-state transitions, differences in total energy between spin states might be extremely sensitive to the approximations adopted in describing the multiplet structure.\cite{MF,takahashi}

The value of $U_{\rm avg}$ does not affect the energy differences between spin-states in the atomic limit, but is relevant for the insulator to metal transition and the gap, as well as for super-exchange. 
Theoretical estimates based on density-functional theory\cite{Hsu,Eder,korotin} yield $\sim 8$~eV in LaCoO$_3$, and we find similar
values with constrained LDA. 
Experimental estimates based on electron spectroscopy\cite{chainiani} yield sizably smaller values ($\sim 3.5$~eV);  
similar values were adopted in many works.\cite{knizek1,knizek2,pandey}
A configuration-interaction cluster model analysis of spectroscopic data yields $U_{\rm avg}=5.5$~eV
for the low and intermediate spin-state and $\sim6.8$~eV for the high-spin state.
Since the value of $U_{\rm avg}$ affects energy scales such as the size of the gap and super-exchange,
when necessary, in the next sections, we present results for several $U_{\rm avg}$.

In conclusion,  our results show that, in the atomic limit,  for all systems analyzed,  a LS$\to$IS scenario is unlikely. At close inspection, LDA+$U$ total-energy calculations are in line with this conclusion,\cite{korotin,zhuang,knizek1,knizek2} %
as they indicate clearly that the stabilization of a given local magnetic state (IS or HS) strongly depends on the number and type of surrounding magnetic neighbors, suggesting that co-operative, band, or at least multi-site effects play a crucial role; at low concentrations, in the absence of magnetic neighbors, the HS appears to be the favored magnetic state.\cite{goodenough,knizek2}
Another important aspect is the evolution of the parameters with increasing temperature.
For LaCoO$_3$ we find that the crystal-field $\Delta_{\rm avg}$ is reduced by about 150~meV
when the temperature increases from 5~K to 1000~K. 
{Remarkably, magnetic susceptibility data\cite{noguchi} indicate that the LS$\to$ HS activation energy, $\Delta_{\rm AE}$, increases with temperature; similarly, the analysis of XAS data by means of the truncated configuration-interaction cluster approach\cite{haverkort} suggests a rise of about 60 meV from 50~K to 700~K.\cite{noguchi}
This result cannot be explained in the atomic limit, even including the effects of spin-orbit interaction, $\lambda {\bf S}\cdot {\bf L}$; we calculate the 
spin-orbit coupling $\lambda$ for LaCoO$_3$ and find $\lambda\sim 54$~meV, slightly larger than the one assumed in model calculations,
\cite{radwanski} but too small to affect the trends on $\Delta_{\rm avg}$. Our results also exclude  that the rhombohedral distortions
increase enough with temperature to overcome the decrease in $\Delta_{\rm avg}$ due to the increasing volume.\cite{rhombohedral}}
\subsection{Super exchange} 
Away from the atomic limit, super-exchange affects the energy balance. However, is the super-exchange energy gain large enough to
be relevant in the spin-state crossover at $T_{\rm SS}$? 
In this section we calculate the super-exchange energy gain for  several types of Co-Co bonds: LS-LS, HS-HS and LS-HS pairs; we do not consider IS-LS pairs, whose formation at low temperature is unlikely.\cite{knizek2}
While differences in the energy of multiplets with fixed number of electrons, relevant to determine the spin-state, are of the order ${\cal{J}}$, virtual excitations of electrons to neighboring sites, relevant for super-exchange, involve the direct Coulomb energy ${\cal{U}}$. Thus Coulomb exchange anisotropy $\Delta {\cal {J}}_{\rm avg}$ has a small effect on super-exchange; for simplicity we neglect it and consider density-density terms only. For a LS-LS pair we obtain the energy gain 
\begin{eqnarray}
\Delta E_{\rm SE}^{\rm LS\mbox{\rm-}LS}\!\!&=& \!\!-
\sum_{\alpha=1}^3\sum_{\alpha^\prime=4}^5
\frac{|t_{\alpha^\prime,\alpha}^{i,i^\prime}|^2+|t_{\alpha^\prime,\alpha}^{i,i^\prime}|^2}
{U_0-5{\cal{J}}+\Delta_{\alpha^\prime,\alpha} },
 \end{eqnarray}
where $\Delta_{\alpha^\prime,\alpha}=\varepsilon_{\alpha^\prime}-\varepsilon_\alpha$.
For a HS-HS pair, assuming that Co ions are in a paramagnetic state, we obtain
 \begin{eqnarray}\nonumber
  \Delta E_{\rm SE}^{\rm HS\mbox{\rm-}HS}\!&=&\!
   -\frac{1}{2n_{\beta}^2}
   \sum_{\{\beta\},\{\beta^\prime\}} \left[ \sum_{\alpha \ne \beta}\sum_{\alpha^\prime\ne\beta^\prime}
   \frac{|t_{\alpha,\alpha^\prime}^{i,i^\prime}|^2+|t_{\alpha^\prime,\alpha}^{i,i^\prime}|^2} 
        {U_0+3{\cal{J}}+\Delta_{\alpha^\prime,\alpha}}
   \right. \\ \nonumber \! 
  &+&   \left.    \! \!\!   \sum_{\alpha^\prime\ne \beta^\prime} 
  \frac{|t_{\beta,\alpha^\prime}^{i,i^\prime}|^2+|t_{\alpha^\prime,\beta}^{i,i^\prime}|^2} {U_0-3{\cal{J}}+\Delta_{\alpha^\prime,\beta}} 
    + \frac{|t_{\beta,\alpha^\prime}^{i,i^\prime}|^2+|t_{\alpha^\prime,\beta}^{i,i^\prime}|^2} {U_0+{\cal{J}} +\Delta_{\alpha^\prime,\beta}}
    \right],
\end{eqnarray}
where $\{\beta\}$ are the $n_{\beta}$ degenerate $t_{2g}$ levels, over which we average.
\begin{figure}
\center
\rotatebox {0}{\includegraphics [width=0.5\textwidth]{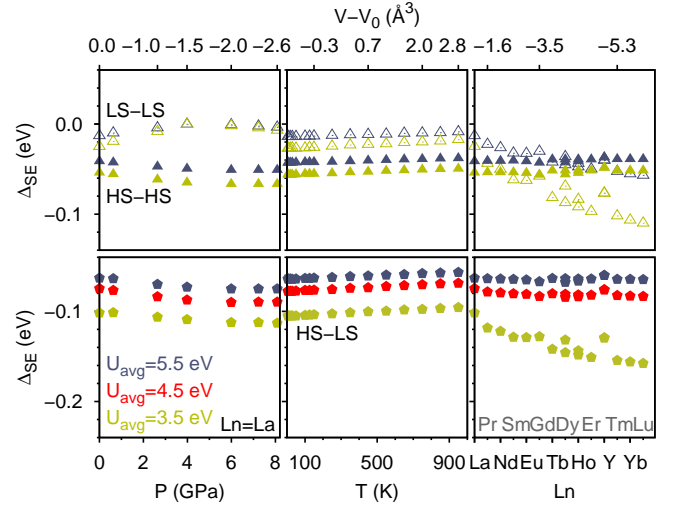}}
\caption{\label{SE}
(Color on-line) 
Super-exchange energy gain per bond, $\Delta_{\rm SE}$, for a LS-LS pair (empty triangles), a HS-HS pair (filled triangles) and a LS-HS pair (pentagons).
Left: $\Delta_{\rm SE}$ as a function of pressure. Center: $\Delta_{\rm SE}$ as a function of temperature.
Right: $\Delta_{\rm SE}$ as a function of Ln ion. 
The corresponding volumes are also given; $V_0$ is the volume of LaCoO$_3$ at room temperature and ambient pressure.
Multiple points for the same compound are results from different structural data sets.}
\end{figure}
Results for material-dependent hopping integrals are shown in Fig.~\ref{SE}. 
We find that for $U_{\rm avg}\sim 5$~eV or larger the energy gains do not change much with increasing ${\cal{J}}$ from 0 to 1 eV.\cite{SEdependenceonJ} 
Instead, for fixed  ${\cal{J}}\sim 0.89$~eV, we find a remarkable change in 
$\Delta E_{\rm SE}^{\rm LS\mbox{\rm-}LS}$
reducing $U_{\rm avg}$  to $\sim 3.5$~eV, because the denominator sizably decrease.
In Fig.~\ref{SE} we show two significant parameter ranges.
For $U_{\rm avg}\sim 5.5$~eV, super-exchange favors a high-spin ground state for LaCoO$_3$ for all temperature
and pressures considered.  In the series LnCoO$_3$, the super-exchange energy gain becomes larger for a LS ground state for ions smaller than Dy$^{3+}$. 
For $U_{\rm avg}\sim 3.5$~eV the crossing happens already for Ln=Sm, and the energy gain per LS-LS bond increases sizably to 90 meV around Ln=Y.

It is at this point crucial to evaluate also the super-exchange energy gain associated with the formation of a HS-LS bond, given by
\begin{eqnarray}
\Delta E^{\rm HS\mbox{\rm-}LS}_{\rm SE}
&=&
\nonumber -\frac{1}{2 n_{\beta}} \sum_{\{\beta\}} 
\left[\sum_{\alpha\ne \beta} 
\sum_{\alpha^\prime=4}^5 
\frac{ |t_{\alpha,\alpha^\prime}^{i,i^\prime}|^2+|t_{\alpha^\prime,\alpha}^{i,i^\prime}|^2} {U_0-{\cal{J}}+\Delta_{\alpha^\prime,\alpha}} \right.
\\   \nonumber
\!\!&+&\!\!
\sum_{\alpha^\prime=4}^5 \left(
\frac{ |t_{\beta,\alpha^\prime}^{i,i^\prime}|^2\!+\!|t_{\alpha^\prime,\beta}^{i,i^\prime}|^2} {U_0\!-\!7{\cal{J}}+\Delta_{\alpha^\prime,\beta}} 
\!+\!
\frac{ |t_{\beta,\alpha^\prime}^{i,i^\prime}|^2\!+\!|t_{\alpha^\prime,\beta}^{i,i^\prime}|^2} {U_0\!-\!3{\cal{J}}+\Delta_{\alpha^\prime,\beta}} \right) \\
&+&\!\! \!\left.
\sum_{\alpha^\prime\ne \beta} \sum_{\alpha=1}^3 
\frac{ |t_{\alpha,\alpha^\prime}^{i,i^\prime}|^2+|t_{\alpha^\prime,\alpha}^{i,i^\prime}|^2} {U_0-{\cal{J}}+\Delta_{\alpha^\prime,\alpha}}
\right].
\end{eqnarray}
\begin{figure}
\center
\rotatebox {0}{\includegraphics [width=0.48\textwidth]{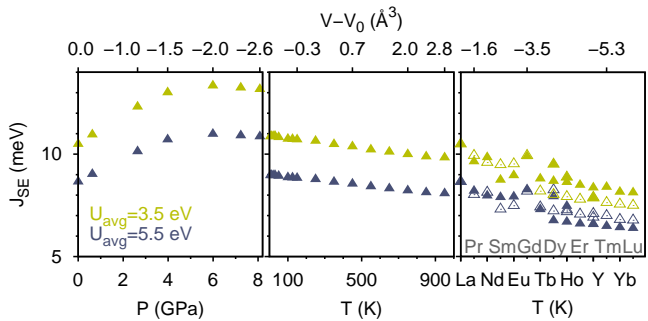}}
\caption{\label{J}
(Color on-line) 
Magnetic coupling $J_{\rm SE}$ for a HS-HS pair, as a function of pressure (left), temperature (center), and of Ln ion (right; empty symbols: (001) direction; filled symbols: (100) or (010) direction). 
The corresponding volumes are also given; $V_0$ is the volume of LaCoO$_3$ at room temperature and ambient pressure.
Multiple points for the same compound are results from different structural data sets.}
\end{figure}
This energy gain is relevant in many scenarios of spin-state crossover.
We find that  $\Delta E^{\rm HS\mbox{\rm-}LS}_{\rm SE}/\Delta E^{\rm HS\mbox{\rm-}HS}_{\rm SE}$ is $\sim 1.5$-$2$  for $U_{\rm avg}\sim 5.5$, and it increases with decreasing $U_{\rm avg}$.
Furthermore, all super-exchange energy gains calculated here vary slowly with increasing temperature.
These results show that, even neglecting lattice relaxation effects,\cite{goodenough} it is energetically favorable to form isolated HS ions rather than HS-HS bonds. 

The super-exchange energy difference between a HS-HS and a LS-LS pair is
$$
\delta E^{\rm LH}=(\Delta^{\rm HS\mbox{\rm -}HS}_{\rm SE}-\Delta^{\rm LS\mbox{\rm -}LS}_{\rm SE}).
$$
For $U_{\rm avg}=5.5$~eV,  we find $\delta E^{\rm LH}<0$ (energy gain if a HS-HS is formed) for systems with larger ionic radius and $\delta E^{\rm LH}>0$ (energy loss) for systems with smaller ionic radius.
The energy cost of a HS-HS pair with respect to two isolated HS ions is
$$
\delta\Delta_{\rm SE} \sim
\Delta E_{\rm SE}^{\rm HS\mbox{\rm-}HS}+\Delta E_{\rm SE}^{\rm LS\mbox{\rm-}LS}
-2\Delta E_{\rm SE}^{\rm HS\mbox{\rm-}LS}.
$$
We find that $\delta\Delta_{\rm SE}>0$ for all systems and all parameter ranges. 
For $U_{\rm avg}=5.5$~eV, it decreases from 70 to 30 meV along the series;
for fixed Ln, it increases with decreasing $U_{\rm avg}$.
Based on these results, we can build a model which describes the system.
We introduce pseudospin operators $\sigma_z$, and identify LS sites with spin down and HS sites with spin up.\cite{IS} The interactions between spins yield a Ising-like model
\begin{eqnarray*}
H_{\rm eff}&=&\sum_i h(T) \sigma_z^i +\frac{1}{2}\sum_{\langle ji\rangle}J \sigma_z^i\sigma_z^j. 
\end{eqnarray*}
Here $n_i=n_i^H+n^L_i=1$ is the occupation per site, $\sigma_z^i=n_i^H-n^L_i$, and $\langle ji \rangle$ are near neighboring lattice sites.
The parameter
$$h(T)=\frac{1}{2}E^{\rm LH}(T)+\frac{1}{4}q\delta E^{\rm LH},$$ 
where $q$ is the coordination number, plays the role of an external field. The temperature enters explicitly only in
\begin{eqnarray*}
E^{\rm LH}(T)&=&-3{\cal J}_{\rm avg}+30\Delta {\cal J}_{\rm avg} +2\Delta^{\rm LH}(T)
\end{eqnarray*}
The coupling is $J=\frac{1}{4}\delta\Delta_{\rm SE}>0$ (antiferro). 
In static mean-field a LS homogeneous state is given by the solution of the self-consistent equation
$$
\langle\sigma_z^i\rangle=\tanh (-h(T) - q J\langle\sigma_z^i\rangle)\beta.
$$
At low temperature $h (T)>0$ is large and the system is in a fully polarized ferro LS state.
Increasing the temperature, $h(T)$ decreases linearly (Fig.~\ref{delta}) allowing the formation of some HS sites. We find that $\delta E^{\rm LH}$ increases with decreasing ionic radius; the crystal-field  is instead maximum around Ln=Eu (Fig.~\ref{delta}).
An increase of activation energy from 1200~K in LaCoO$_3$ to 3200~K in
EuCoO$_3$ has been indeed reported.\cite{recoo2}

When HS states form, covalency-driven lattice relaxation effects\cite{goodenough,haverkort}   
change the environment. In our model, if we take into account only two-sites interactions, this effect enhances $J\to J^\prime=J+J_{\rm rel}$, thus favoring a AF HS/LS phase, described in mean field by
$$
\langle\sigma_z^i\rangle=\tanh (-h (T)+ q J^\prime\langle\sigma_z^i\rangle)\beta.
$$
An overestimate of $J_{\rm rel}$ can be obtained by using the difference between HS and LS in empirical ionic radii ($\delta r_{\rm IR} \sim 0.06\AA$) and the fact that $\Delta$ varies linearly when changing the Co-O distance by a small amount; we can estimate the slope from our results (Fig.~\ref{delta}). 
We find $J_{\rm rel}\sim \frac{1}{4}2\delta \Delta$, with $2\delta \Delta\sim \frac{1}{q}2\Delta^\prime \delta r_{\rm IR}\sim 90$ meV, decreasing to $80$~meV at high temperatures, i.e.\ comparable with  the super-exchange term $\delta \Delta_{\rm SE}$.

If $h(T)=0$, eventually an antiferro ordered HS+LS lattice forms; the critical temperature for such a state can be estimated using mean-field theory, $k_BT_{HS+LS}= qJ^\prime$.
If  $J_{\rm rel}=0$, for $U_{\rm avg}=5.5$~eV $T_{HS+LS}\sim1200$~K for Ln=La, decreasing to 500~K for Ln=Lu; $J_{\rm rel}$ enhances $T_{HS+LS}$ of about 1300~K. 

The critical temperatures we obtain are very large. However, although $h(T)\to0$ with increasing temperature, it likely remains comparable to $J^\prime$ till very high temperatures; thus the lattice stays disordered, and perhaps even close to
a LS ferro solution in a large temperature range, with
$\langle\sigma_z^i\rangle$ (i.e., the occupation of LS states) decreasing linearly with increasing temperature. 
Indeed, in LaCoO$_3$ it has been reported that the fraction of HS sites increasing slowly from
0.1 to 0.4 increasing the temperature from 100 to 700~K.\cite{haverkort}
Hysteresis effects could arise because of the lattice relaxation. 

\begin{figure}[t]
\center
\rotatebox {270}{\includegraphics [width=.6\textwidth]{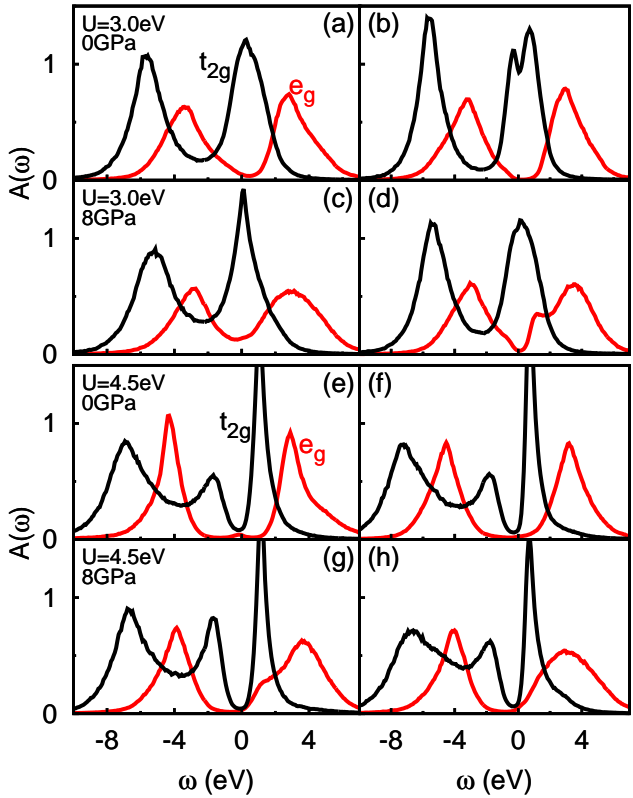}}
\caption{(Color on-line) \label{spectra}
Orbital-resolved LDA+DMFT spectral functions ($t_{2g}$ and $e_g$) of LaCoO$_3$ for decreasing temperature, approaching the insulator to metal transition from above  ($T/T_{\rm IM}\sim 4 \to 2 $).
Upper panels:   $U=U_{\rm avg}=3$~eV, P=0 GPa and 8 GPa. 
Lower panels: $U=U_{\rm avg}=4.5$~eV, P=0 GPa and 8 GPa.}
\end{figure}

Up to here we have assumed that in HS-HS pairs Co HS ions stay paramagnetic and paraorbital.
However, a specific molecular state could form as well as a spin singlet or triplet.
We find that orbitals in the degenerate $t_{2g}$ states play a minor role, because the energy differences of different orbital states with respect to the orbital average is small.
Instead, the formation of a magnetic singlet could play a role. In order to estimate the associated energy gain, we calculated the coupling constant $J_{\rm SE}$ of the Heisenberg interaction $J_{\rm SE}\,{\bf S}_i\cdot {\bf S}_j$ for a HS pair.
We find that  $J_{\rm SE}$ is AFM for all systems (see Fig.~\ref{J}). 
In LaCoO$_3$, for $U_{\rm avg}=5.5$~eV we find $J_{\rm SE}\sim 7$~meV at room temperature and ambient pressure;
it increases to 10~meV  if the pressure rises to 8~GPa, and slightly decreases with  increasing temperature.
Reducing the ionic radius the anisotropy of $J_{\rm SE}$ increases, but the average magnetic coupling decreases down to 8~meV for Ln=Lu.  
Next we assume that in all system, due to the spin-orbit coupling, the effective magnetic moment $p=\sqrt{J(J+1)}\sim \sqrt{2}$; then, the magnetic energy gain associated with the formation of a singlet is $-\frac{9}{4}J_{\rm SE}$, which in LaCoO$_3$ is $\sim -18$~meV. This energy gain partially  reduces $\delta\Delta_{\rm SE}$ and therefore $J$, which would change sign around the end of the series if $U_{\rm avg}\sim 5.5$~eV;
$\delta \Delta_{\rm SE}$ increases however rapidly with decreasing $U_{\rm avg}$. 
The smaller $T_{\rm HS+LS}$ the more favorable is a disordered phase, reinforcing the conclusions of the paragraph above.

However, the magnetic coupling $J_{\rm SE}$ is sizable, and magnetic interactions would be crucial in a homogeneous HS-HS state.
In mean-field theory, the critical temperature $T_N\sim \frac{9}{4} J_{\rm SE}\, J(J+1)$ should then be $\sim 
500$~K for LaCoO$_3$, decreasing in the series to 400~K 
and increasing under pressure to about 700~K. 
The mean-field values are overestimates;
quantum-fluctuation effects, further crystal-field splittings reducing the effective magnetic moment, and the  zero-field splitting,  reduce $T_N$. Still, even taking reduction factors into account, many cobaltates, if in a homogeneous HS state, should be magnetic because the spin-state transition occurs at $T_{\rm SS}<T_N$; in the presence of lonely HS-HS pairs, anti-ferromagnetic short range correlations could be detected. 
On the other hand, if the number of ions thermally excited to a HS state is small 
till high temperature and the phase is disordered, weak FM short-range correlations, as reported in dynamic neutron-scattering experiments, could perhaps be triggered 
{by a double-exchange like mechanism,\cite{Eder}}
or even alone by the small but finite zero-field splitting.\cite{noguchi} 

\begin{figure}[t]
\center
\rotatebox {270}{\includegraphics [width=.32\textwidth]{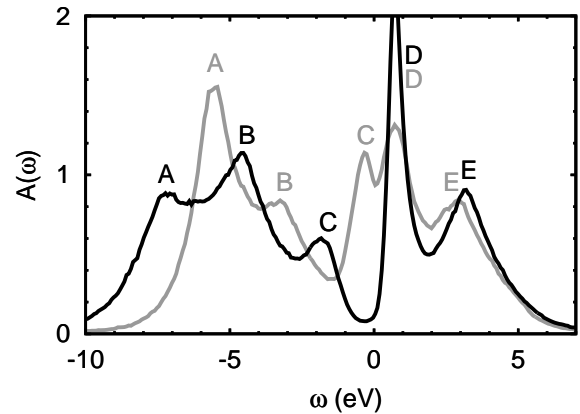}}
\caption{ \label{tot}
Total LDA+DMFT spectral function ($T/T_{\rm IM}\sim 3$) for LaCoO$_3$; $U_{\rm avg}=3$~eV (light) and
$U_{\rm avg}=4.5$~eV (dark).
}
\end{figure}
\subsection{High-temperature phase}

The transition observed at $T_{\rm IM}$ is usually ascribed to a semiconductor-to metal transition, the nature of which is hotly debated. There are indications that strong correlations play a crucial role, and that the spin-state transition is also a crossover from a spin-gapped insulator to a Mott insulator.\cite{tokura,yamaguchi} 
In the absence of firm evidence of the formation of superlattices, we analyze to what extent the insulator to metal transition can be understood in terms of a Mott transition within the homogeneous phase. Several works indeed suggest that the high-temperature phase could be a metallic homogeneous IS state. \cite{knizek2,kiomen}
To clarify if this can be the case, we solve the Hamiltonian (\ref{H}) by means of the LDA+DMFT approach;\cite{details} 
we perform calculations for LaCoO$_3$ (Figs.~\ref{spectra}, and \ref{tot}), for which spectral properties are best known.
The experimental gap is $\sim 0.2$~eV at low temperatures and slightly larger ($\sim 0.5$~eV) above 300~K;
to be consistent with experimental gaps, we vary $U_{\rm avg}$ between 3 eV and 5 eV.

Fig.~\ref{spectra} shows the LDA+DMFT results for $U_{\rm avg}=3$~eV and the ambient pressure structure 
for temperatures approaching $T_{\rm IM}$ (panels (a) and (b)). We find a homogeneous high-spin state; 
thus we do not find support for the proposed\cite{knizek2,kiomen} homogeneous IS state at high-temperature.
Remarkably, we find metallic $t_{2g}$ and insulating $e_g$ spectral-functions, i.e., an {\it orbital-selective} Mott state.
This can be understood from the fact the $t_{2g}$ electrons have a larger orbital degeneracy,\cite{koch} although a smaller band-width (Fig.\ref{W}),
while the $e_g$ electrons are half-filled in the HS state, and therefore the exchange coupling ${\cal{J}}_{\rm avg}$ effectively
enhances the Coulomb repulsion. Decreasing the temperature towards $T_{\rm IM}$ a pseudo gap opens in the $e_g$
spectral-function, and eventually an insulating regime with a small gap and a corresponding bad metal behavior appears. 
We find that the multiplet positions in the spectral function are close to the atomic limit high-spin curve in Fig.~\ref{atomic}.
We repeat the calculation for the 8 GPa structure (Fig.~\ref{spectra}, panels (c) and (d)); this system has a larger crystal-field $\Delta_{\rm avg}$, very close, for the chosen exchange parameters, to the low spin to high-spin ground-state crossover in the atomic limit.
Compared to the ambient pressure case, the $e_{g}$ gap is smaller, while, at fixed temperature, the $t_{2g}$ spectral function remains metallic with higher low-energy density of states. 
These results indicate that, as far as the state remains HS, at 8 GPa the insulator to metal transition should occurr at lower temperature than at ambient pressure, i.e.\ the system  should stay metallic. Experiments show however that LaCoO$_3$ returns to the insulating state under pressure.\cite{pressure2,pressure4}
Our results exclude that this can happen in a HS scenario, and are instead compatible with the suggestion\cite{pressure2} that the metal-insulator transition observed under pressure is driven by a spin-state transition. 

Increasing $U_{\rm avg}$ to $\sim 4.5$~eV we find (Fig.~\ref{spectra}, panels (e), (f), (g), (h)) that a real gap opens in the $t_{2g}$ spectral function, and both $e_g$ and $t_{2g}$ spectral functions are insulating even at high temperatures. The $t_{2g}$ gap is small, still  compatible with the bad metal behavior observed for  $T>T_{\rm IM}$. 
Furthermore, calculations for the high-pressure structure show that the $t_{2g}$ spectral function does not change much while the $e_g$ gap is reduced. {{This can be understood observing that the $e_g$ band-width increases substantially more that the $t_{2g}$ band-width increasing pressure
(Fig.~\ref{W}).} %

\begin{figure}[t]
\center
{\rotatebox {0}{\includegraphics [width=.47\textwidth]{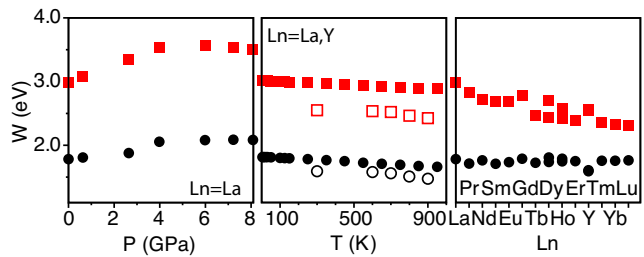}}}
\caption{(Color on-line) \label{W}
LnCoO$_3$ series: Evolution of the $e_g$ (squares) and $t_{2g}$ (circles) bandwidth 
with pressure (Ln=La, R$\bar{3}$c structure), temperature (filled symbols: Ln=La (R$\bar{3}$c structure) and
empty symbols: Ln=Y), and ionic radius Ln$^{3+}$ (right panel).}
\end{figure}
The total spectral function at ambient pressure is shown in Fig.~\ref{tot}.
We find three negative energy peaks (A,B,C), and two positive energy features (D,E).
For $U_{\rm avg}\sim 4.5$~eV peak A is at $-7.5$~eV, peak B at $-4.7$~eV,  $\sim 2$~eV above, and peak C at $\sim -1.9$~eV, features all observed in XPS, UPS, or photoemission data.\cite{chainiani,saitoh,pandey,pes} 
The spectra for $U_{\rm avg}$=3~eV and $U_{\rm avg}$=4.5 eV differ in particular in the photoemission part;
reducing $U_{\rm avg}$ to 3~eV, we find that the spectrum is moved almost rigidly toward the right by about 2~eV; the first and second features move to -5.5~eV and -3.5~eV respectively, while the lowest energy peak moves to $-0.5$~eV, very close to the Fermi edge, to partially merge with the $0.8$~eV peak; finally, spectral weight moves from B to C and A.
While the exact positions of peaks A, B, and C shift with $U_{\rm avg}$, the overall shape of the spectral function appears in line
with XPS, BIS, XAS and PES data at room temperature.\cite{medarde,pes,chainiani}
The positive energy features D and E at $\sim1$~eV and $\sim 3$~eV are  reflected in the form of the XAS and BIS spectra.\cite{chainiani}

Thus the insulator to metal transition, as described  in a homogeneous scenario, and the spin-state crossover  
exhibit different trends;
the spin-state crossover is controlled by small changes in $\Delta$, exchange anisotropy and super-exchange, parameters
which change sizably with decreasing ionic radius. 
If the homogeneous HS is populated, a Mott insulator to bad metal transition can occur, at a  temperature which depends strongly on the $t_{2g}$ band-width and crystal-fields.

\section{Conclusions}
We have studied the nature of the spin-state and metal-insulator transition in LnCoO$_3$ cobaltates.
We show that a low-temperature intermediate spin-state scenario is unlikely in the full LnCoO$_3$ series.
We show that the spin-state transition is controlled not only by the cubic
crystal-field $\Delta$ and the average Coulomb exchange ${\cal{J}_{\rm avg}}$ but, surprisingly, 
also by the Coulomb exchange anisotropy $\Delta {\cal{J}_{\rm avg}}$  and by super-exchange. 
We propose a simple Ising-like model to describe the spin-state transition. 
We find that lattice relaxation and super-exchange yield anti-ferro coupling of the same order, which compete with
the crystal field. Our model qualitatively explains the trends observed in spin-state transitions in the LnCoO$_3$ series.
By using the LDA+DMFT approach, we show that in LaCoO$_3$ a high-temperature homogeneous intermediate spin state is unlikely.
Within a homogeneous HS state, we find that the HS metal-insulator transition has a different nature than the spin-state transition, as it mostly depends on the $t_{2g}$ states and it {could} be orbital-selective. %

\section{Acknowledgements}
We acknowledge discussions with H.~Tjeng. Calculations were done on the J\"ulich Blue Gene/P and Juropa. We acknowledge financial support from the Deutsche Forschungsgemeinschaft through research unit FOR1346.


\begin{thebibliography}{99}
\bibitem{almostfull} In order to reduce the computational effort, most calculations are performed keeping only density-density terms; we find that differences due to the neglected terms are small in the relevant parameter regime.
\bibitem{noteonfelectrons}
We find that  the parameters depend very weakly on the position of RE $4f$ levels;
we obtain very similar results if we treat $f$ electrons within LDA+$U$ ($U\in[0,7]$~eV),
or if we replace RE$^{3+}$ with La$^{3+}$ (empty $4f$ shell).
\bibitem{nmtocef}
The crystal-field splittings obtained with the NMTO approach are slightly larger, but the trends remain the same. 
In the low-temperature ambient pressure phases $\Delta\sim 2.15$~eV
while  $\Delta\sim 2.4$~eV under a pressure of 8 GPa and  $\sim 2.05$~eV at 900~K.
For YCoO$_3$ we find $\sim 2.15$~eV at 300~K and $2.0$ at $\sim 900$~K.
The change in crystal-field splitting between high-spin and low spin phases is in all cases at most 300 meV.
\bibitem{atomic} For 3$d$ elements, the ratio $F_4/F_2$ is usually approximated to $\sim 0.625$; this yields
$\Delta {\cal{J}}_{\rm avg}/ {\cal{J}}_{\rm avg} \sim 0.08 $.
The extreme values for $\Delta {\cal{J}}_{\rm avg}/ {\cal{J}}_{\rm avg}$ are 0 ($F_4=F_2$) and 0.2 ($F_4=0$).
\bibitem{SEdependenceonJ}
We find that, for ${\cal{J}}_{\rm avg}$ between 0 and 1 eV, the super-exchange energy gain 
is reduced of about 30 meV for $S=2$ and increased of about the same amount for $S=0$;
the trends remain the same.
\bibitem{IS} The IS state could be included in our model as a third excited state. In this case
instead of a  Ising model for a pseudospin 1/2, we would obtain an Ising model for a pseudospin 1.
\bibitem{details} To reduce the sign problem we dropped Coulomb terms without a density-density matrix form; this moves the spin-state transition to slightly larger values of $\Delta_{\rm LH}$ ($2.8{\cal{J}}_{\rm avg}$ versus $2.6 {\cal{J}}_{\rm avg}$), as discussed in section II A, but do not affects the conclusions.
\end{thebibliography}
\end{document}